\documentstyle[preprint,aps]{revtex}

\begin{document}

\bibliographystyle{prsty}

\title{On the physical interpretation of effective actions using
Schwinger's formula}

\author{L. C. de Albuquerque\thanks{e-mail: farina@vms1.nce.ufrj.br},
C. Farina, Silvio J. Rabello and  Arvind N. Vaidya}

\address{\it Instituto de F\'\i sica, Universidade Federal do Rio de
Janeiro, \\ Rio de Janeiro,  RJ, Caixa Postal 68.528-CEP 21945-970, Brasil}

\date{\today}

\maketitle

\begin{abstract}

{\sl We show explicitly that Schwinger's formula for one-loop effective
actions corresponds to the summation of energies associated with the
zero-point oscillations of the fields. We begin with a formal proof, and
after that we confirm it using a regularization prescription.}

\end{abstract}


An explanation for the Casimir energy \cite{cas} is that it appears when
one distorts the vacuum of a quantum field theory by introducing boundaries
or background fields. One way to study this vacuum distortion is to
perform the canonical quantization of the theory and the result is that this
energy turns out to be the sum of zero-point oscillations of the quantum
fields. In the functional approach the Casimir energy emerges as an
effective potential induced by the vacuum fluctuations in the presence of
these external perturbations \cite{DR}.

The connection between these two points of view was explored in \cite{CBern}
and more recently studied in \cite{Myers} with the aid of the
$\zeta$-function method to evaluate the effective potential and turn it
into a sum of zero-point energies.

In this paper we use the proper time formula derived long ago by Schwinger,
in his early investigations on effective actions in QED \cite{schw2}, and
recently applied by himself to the Casimir effect \cite{schw1},
to establish the connection of the effective potential with the zero-point
oscillations. This formula is widely used in quantum field theory, in special
when one is confronted with background fields as in the semiclassical
approach to quantum gravity \cite{BD}. We first employ this formula
(hereafter Schwinger's formula) in a rather formal way with regard to the
divergences that appear during the calculations and then proceed in a
regularized fashion, obtaining a regularized sum of zero-point energies.

Although the result we are going to prove is quite general, we choose for
simplicity a definite problem in order to emphasize the basic points of
our demonstration without dealing with involved mathematical details.

Consider a massive scalar field in a $(3+1)$-dimensional space-time
(the extension to $d+1$ dimensions is trivial) constrained by two
flat plates of area A, orthogonal to the z axis and placed at $z=0$ and
$z=a$ (we assume that $A\gg a^2$). The boundary conditions on the plates
are chosen to be of the Dirichlet type so that the fields must vanish at
$z=0$ and $z=a$.

In a recent paper \cite{schw1} this energy was obtained, for the massless
case, using the Schwinger's formula

\begin{equation}
\label{sf}
W_0=-{i\over2}Tr\int_{s_0}^\infty {ds\over s} e^{-is(H
- i\epsilon)}\,,
\end{equation}
where $ W_0$ is the one-loop effective action and $H$ is the
proper-time Hamiltonian for the system at hand. The symbol $Tr$ refers to
the trace over all the degrees of freedom, including those of spacetime.

Then, the Casimir energy  is simply given by ${\cal E}=-{ W_0\over T}$,
where $T$ can be viewed as a time interval. In equation (\ref{sf}), the
regularizing cut-off $s_0$ must be put to zero only after a suitable
subtraction of non-physical terms is made \footnote{An application of
this formula for the same problem but with a different regularization
prescription can be found in \cite{far1}. See also \cite{far2} for the
massive case.}.

We first establish a formal connection and after that, we confirm our
result in a more rigorous way, using a regularization procedure. Let us
start with the formal expression with $s_0=0$:

\begin{equation}
\label{2}
 W_0=-{i\over2}Tr\int_0^\infty {ds\over s}  e^{-is( H
- i\epsilon)}\,,
\end{equation}
where $H={\bf p}^2-\omega^2+m^2$, with ${\bf p}=-i{\bf\nabla},\;
\omega=i\partial_t$. Taking a derivative with respect to $m^2$ on both sides
of (\ref{2}), performing the trace and making the identification
${\cal E}=- W_0/T$, we have

\begin{equation}
\label{4}
{1\over A}{\partial{\cal E}\over \partial m^2}=
{1\over2}\sum_{n=1}^{\infty}\int{dk_1dk_2dk_0\over (2\pi)^3}
\int_0^{\infty}ds e^{-is(-k_0^2+k_1^2+k_2^2 +{n^2\pi^2\over a^2} -i\epsilon
+m^2)}\,.
\end{equation}
Integration over $s$ immediately leads to

\begin{equation}
\label{5}
{1\over A}{\partial{\cal E}\over \partial m^2}=
{i\over2}\sum_{n=1}^{\infty}\int\int{dk_1dk_2\over (2\pi)^2}
\int{dk_0\over 2\pi}
{1\over {k_0^2-[k_1^2+k_2^2 +{n^2\pi^2\over a^2}-i\epsilon +m^2]}}\,.
\end{equation}
Now, with the aid of the residua theorem, we can integrate over $k_0$ to get

\begin{equation}
\label{6}
{1\over A}{\partial{\cal E}\over \partial m^2}=
{1\over2}\sum_{n=1}^{\infty}\int\int{dk_1dk_2\over (2\pi)^2}
{1\over \sqrt{k_1^2+k_2^2 +{n^2\pi^2\over a^2} +m^2}}\,.
\end{equation}
Then, making the integration over $m^2$ we obtain,

\begin{equation}
\label{7}
{{\cal E}\over A}=
\sum_{n=1}^{\infty}\int\int{dk_1dk_2\over (2\pi)^2}
{1\over2}\omega(k_1,k_2,n)\,,
\end{equation}
apart from an additive irrelevant constant, and where we defined the
eigenfrequencies (zero-point energies of the field)
$\omega(k_1,k_2,n)=\sqrt{k_1^2+k_2^2 +{n\pi\over a}^2 +m^2}$.
Expression (\ref{7}) is precisely the usual (unregularized) mode summation
expression for the Casimir energy of a massive scalar field between two
parallel plates distant apart a distance $a$ \cite{phys.rep.}.

Since in our previous deduction manipulations with divergent terms were made,
let us now obtain the same result in a more rigorous way,
what we mean is that we shall establish a connection between two
regularized expressions, Schwinger's one and a (regularized) mode
summation expression, with some kind of regularization prescription adopted.

However, instead of starting with the regularized expression ({\ref{sf})
(with the cut-off $s_0$), we adopt the following regularization prescription

\begin{equation}
\label{8}
 W_0=-{i\over2}Tr\int_0^\infty ds s^{\nu-1}  e^{-is(H
-i\epsilon)}\,,
\end{equation}
where $\nu$ is large enough to make the integral well defined. In this
approach, after the integral is computed an analytical continuation to the
whole complex plane of $\nu$ must be done and then the limit
$\nu\rightarrow 0$ must be carefully taken (sometimes an appropriate
subtraction must be made, see \cite{far1,far2} for more details on
this approach)

As before we start by taking the derivative of (\ref{8}) with respect to
$m^2$. After evaluating the trace, we get

\begin{equation}
\label{9}
{\partial W_0\over \partial m^2}=
-{AT\over2}\sum_{n=1}^{\infty}\int{dk_1dk_2dk_0\over (2\pi)^3}
\int_0^{\infty}ds s^{\nu}e^{-is(-k_0^2+k_1^2+k_2^2 +{n^2\pi^2\over a^2}
+m^2)}\,.
\end{equation}
Using the definition of the Euler Gamma function, the integration over $s$
readily yields
\begin{equation}
\label{10}
{\partial W_0\over \partial m^2}=
-(i^{\nu+1}){AT\over2}\Gamma(\nu+1)\sum_{n=1}^{\infty}
\int\int{dk_1dk_2\over (2\pi)^2}\int{dk_0\over 2\pi}
[k_0^2-(k_1^2+k_2^2 +{n^2\pi^2\over a^2}+m^2)]^{-(\nu+1)}\,.
\end{equation}
Using now that
\begin{equation}
\label{int}
\int_{-\infty}^{\infty}dx(x^2+\alpha^2)^{-\sigma}=(\alpha^2)^{{1\over2}-
\sigma} B({1\over2},\sigma-{1\over2})\,,
\end{equation}
where $B(p,q)={\Gamma(p)\Gamma(q)\over \Gamma(p+q)}$, we get
\newpage
\begin{equation}
\label{11}
{\partial W_0\over \partial m^2}=
-(i^{-\nu}){AT\over2}\Gamma({1\over2})\Gamma(\nu+{1\over2})
\sum_{n=1}^{\infty}\int\int{dk_1dk_2\over (2\pi)^2}{1\over 2\pi}
[k_1^2+k_2^2 +{n^2\pi^2\over a^2}+m^2]^{-(\nu+{1\over2})}\,.
\end{equation}
Integrating on $m^2$ and identifying ${\cal E}=-{ W_0\over T}$,
we finally obtain apart from an irrelevant additive constant

\begin{equation}
\label{12}
{{\cal E}\over A}=
\sum_{n=1}^{\infty}\int\int{dk_1dk_2\over (2\pi)^2}
{1\over2}\omega(k_1,k_2,n)F(\nu)\,,
\end{equation}
where we defined

\begin{equation}
\label{13}
F(\nu)={i^{-\nu}\over {1-2\nu}}{\Gamma(\nu+{1\over2})\over \sqrt{\pi}}
\omega^{-2\nu}(k_1,k_2,n)\,.
\end{equation}

Some comments are in order here. Expression (\ref{12}) is nothing but the
usual mode summation with the presence of the regulator function $F(\nu)$,
since $F(\nu)$ contains the negative power $\omega^{-2\nu}$. So, for $\nu$
large enough expression (\ref{12}) is well defined and after an analytical
continuation to the whole complex $\nu$ plane is made, the limit
$\nu\rightarrow0$ can be (carefully) taken to yield the physical Casimir
energy per unit area. It is easily seen that as $\nu\rightarrow0$ we have
that the regulator $F(\nu)\rightarrow 1$. Hence, we have established the
equivalence between the modified Schwinger's formula (\ref{8}) and the
(regularized) mode summation approach. Although we have chosen a definite
problem, i.e. the Casimir energy for a massive scalar field with
Dirichlet boundary condition in one direction, our deduction is general, and
therefore the Schwinger's formula for the (one-loop) effective action can
always be interpreted as a summation over zero-point energies.

One of us (CF) would like to thank M. Asorey, A. J. Segu\'\i-Santonja
and M. V. Cougo Pinto for helpful discussions on this subject. This
work was partially supported by CAPES and CNPq (Brazilian councils of
research).

\end{document}